\documentclass[12pt,a4paper]{article}

\usepackage{graphicx}
\usepackage{cite}
\usepackage{float}
\usepackage{amsmath,amssymb,amsfonts}
\usepackage{bbm}

\newcommand{\bse}{\begin{subequations}}
\newcommand{\ese}{\end{subequations}}
\newcommand{\be}{\begin{equation}}
\newcommand{\ee}{\end{equation}}
\newcommand{\bea}{\begin{eqnarray}}
\newcommand{\eea}{\end{eqnarray}}
\newcommand{\ba}{\begin{array}}
\newcommand{\ea}{\end{array}}

\input amssym.def
\input amssym.tex

\usepackage[colorlinks=true, linkcolor=blue, bookmarks=true, citecolor=red]{hyperref}

\usepackage{geometry}
\begin{document}

\title{From Schwinger Balls to Black Holes}
\date{}

\maketitle

\begin{center}

\hrule
\vspace{5mm}

\author{Davood Allahbakhshi\footnote{allahbakhshi@ipm.ir}\\\vspace{5mm}School of Particles and Accelerators, Institute for Research in Fundamental Sciences (IPM), P.O.Box 19395-5531, Tehran, Iran}

\end{center}

\vspace{5mm}
\hrule

\linespread{1.5}

\abstract{
\noindent
We have shown intriguing similarities between Schwinger balls and black holes. By considering black hole as a gravitational Schwinger ball, we have derived the Bekenstein-Hawking entropy and the first law of black hole thermodynamics as a direct result of the inverse area dependence of the gravitational force. It is also shown that the Planck length is nothing but the gravitational Schwinger length. The relation between the mass and the radius of the black hole is derived by considering the black hole as a Schwinger ball of gravitons. We show how the evolution of the entanglement entropy of the black hole, as Page introduced many years ago, can be obtained by including gravitons in the black hole's evaporation process and using a deformed EPR mechanism. Also this deformed EPR mechanism can solve the information paradox. We show how naive simultaneous usage of Page's argument and equivalence principle leads to firewall problem.}

\newpage

\hrule

\tableofcontents

\vspace{5mm}
\hrule

\section{Introduction}

Black holes, the mysterious solutions of general relativity, have raised many confusions and questions from their discovery in 1915 by Schwarzschild. There are many surprising features associated with them. Black holes exist in pure gravitational theories without any matter and general relativity does not say anything about their origin. The physical meaning of the horizon, as an unreachable barrier from the viewpoint of the static observer is hard to understand while the free falling observer can see nothing. Things got worse when Carter, Hawking and Bardeen found that black holes obey some thermodynamical laws \cite{Bardeen:1973gs}. These thermodynamical laws added two more, problematic features to black holes which are the temperature and entropy proportional to the surface gravity and area of the horizon respectively. The existence of entropy means that black holes should have some microstates. The problem is that the number of these microstates is proportional to the area of the black hole but not its volume like any other system we know. We do not know these mysterious microstates until now. For temperature, situation got better. Very soon Hawking found that the horizon of the black hole radiates a flow of thermally distributed particles with the temperature proportional to the surface gravity of the horizon \cite{Hawking:1974sw}. So the horizon is a hot surface which radiates like a black body with the required temperature. This radiation causes the black hole to be evaporated. But there is another problem. The spectrum of this radiation is thermal while the initial matter can be in a pure state. It means that the total process of black hole formation and evaporation can change a pure state to a mixed state which is not possible by a unitary quantum evolution. If it really happens then the information of the state of the initial matter will be lost. So we have the famous \emph{information paradox} \cite{Hawking:1976ra}.

Many physicists tried to solve the information paradox in many different ways. The first possible resolution comes from the fact that Hawking thermal radiation is the result of a semi-classical calculation for an eternal black hole.  For a real evaporating black hole we need to take into account the backreaction of the radiation on the black hole as well as some quantum effects. Such modifications will make the radiation non-thermal and so not completely informationless. Some physicists believe that this effect is not large enough to solve the problem \cite{Mathur:2009hf}, in contrast to some others \cite{Dvali:2015aja}.

Also more complicated explanations are proposed for solving the paradox. One of these explanations is complementarity \cite{'tHooft:1990fr,Susskind:1993if} which states that the information of the infalling matter will be duplicated. One copy goes into the black hole with the infalling matter and the other copy will be reflected back to the static observer at infinity with radiated particles. An observer has access to just one of these copies and these two pictures are complementary.

Some physicists believe that the paradox can be solved by possible nonlocal effects \cite{Giddings:2012gc}. Also existence of some macroscopic or many microscopic wormholes is proposed to link the inside and outside of the black hole \cite{Frolov:1993jq,Page:1993up,Maldacena:2013xja}.

A very interesting picture of black holes is proposed by Dvali and Gomez \cite{Dvali-Gomez-papers}. In this picture a black hole is a Bose-Einstein condensate of gravitons, stuck at the critical point of a quantum phase transition. This critical point corresponds to a maximally packed ball of soft gravitons. Then by borrowing the relation between the mass and radius of the black hole from general relativity, they show that the occupation number of these soft gravitons obeys the relation
\be\label{Dvali-Gomez}
N=\frac{R^{d-2}}{L_P^{d-2}},
\ee
where $R$ and $L_P$ are the Schwarzschild radius of the black hole and the Planck length and $d$ is the number of space-time dimensions. This very important relation immediately leads to Bekenstein entropy, since the entropy is proportional to $N$.

Similarities between Hawking radiation, Unruh effect \cite{Fulling:1972md,Davies:1974th,Unruh:1976db} and Schwinger effect \cite{Sauter:1931zz,Heisenberg:1935qt,Schwinger:1951nm} are mentioned by some physicists \cite{Parentani:1991tx,Frolov:2014wja,Kim:2016dmm}. Schwinger effect is the instability of the vacuum to create couples of particle-antiparticle in an external electric field. The effect is significant when the electric field is stronger than the Schwinger limit
\be\label{Schwinger-limit}
E_c = \frac{m^2c^3}{e\hbar}.
\ee
The electric field above can be understood as the field which can create a couple of charged particles of mass $m$ in a distance of order of the de Broglie wavelength of the particle
\bea\label{Schwinger-simple}
\left\{
\begin{array}{l}
eE_c\lambda = mc^2 \cr
\lambda = \frac{h}{mc}
\end{array}\right.
\Rightarrow E_c = \frac{m^2c^3}{eh}.
\eea
The spectrum of the created particles is thermal\footnote{In fact the spectrum is thermal in \emph{mass}, if the momentum of the particle is zero, but not in momentum.} with the temperature proportional to $E$. In this sense the \emph{Schwinger temperature} is comparable to \emph{Unruh temperature} which is proportional to $a$, the acceleration of the non-inertial observer.

In this paper we show that the similarity between Schwinger, Unruh and Hawking effects is more than this. In fact we derive the equations and quantities which are the characteristics of the Schwarzschild black hole from a \emph{gravitational Schwinger effect}. This means that from \emph{microscopic} point of view, the black hole is in fact a gravitational Schwinger ball although its classical picture is completely different in general relativity. A very important point here is believing that the picture which general relativity presents for black holes is not completely correct. We need to consider the black hole as a system of gravitons in a special state, not a mysterious geometrical object like what general relativity presents. Of course replacing quantum objects with effective classical geometrical objects has some failures since they are acceptable to some approximation at some limit. Away from this limit, these geometrical objects may lead to some unphysical results.

The paper is organized as follows. In next section we introduce the electric and gravitational Schwinger balls and we calculate some of their main characteristic quantities such as their size and entropy. For gravitational Schwinger ball we show how a small change in the mass of the ball leads to the first law of black hole thermodynamics. In section 3, we explain the evaporation process of the black hole from the microscopic point of view, inspired by electric Schwinger ball. We will see how the entanglement entropy of the black hole increases to a maximum value and how can decrease after that. We will see how the information paradox will be solved in this microscopic picture. It is also explained how the firewall problem arises and will be resolved. Finally we finish the paper by summary and discussion.

\section{Schwinger Balls}
In this section we introduce the electric and gravitational Schwinger balls and demonstrate their similarities.

\subsection{Electric Schwinger Ball}

Suppose that we have a ball of electrons. If this charge ball is small enough (we will see how small it should be), then inside a region around it the electric field is stronger than the Schwinger limit
\be
E_c = \frac{m_e^2c^3}{e\hbar},
\ee
where $e$ and $m_e$ are the charge and the mass of the electron. We name this region the \emph{Schwinger Ball}. RG flow of the electric coupling ensures that this region exists. Inside this Schwinger ball the electric field creates charged particles and anti-particles, namely electrons and positrons. Because of the electric field of the charge (electron) ball, the electrons are pushed away and positrons are attracted to the charge ball. This flow of positrons will annihilate the electrons of the charge ball when they interact with them. Produced photons will (or will not!) escape from the Schwinger ball and the ball will be \emph{evaporated} to electrons and photons.

As the starting point let us calculate the radius of the Schwinger ball that we name \emph{Schwinger Radius}. Semi-classically we just use the Coulomb's law. Of course any modification to Coulomb's law due to quantum effects of QED will modify our results as well. The Schwinger radius is where the electric field around the charge ball equals to the Schwinger limit, so we have
\bea\label{electric-Dvali-Gomez}
&&E_c = \frac{1}{\varepsilon_0}\frac{Q}{R_c^{d-2}} \cr\cr
\Rightarrow &&\frac{m^2c^3}{e\hbar} = \frac{1}{\varepsilon_0}\frac{Q}{R_c^{d-2}} \cr\cr
\Rightarrow && R_c^{d-2} = N_e \; L_{S}^{d-2},
\eea
where $N_e = Q/e$, is the number of electrons inside the Schwinger ball and $L_{S}$ is a chracteristic length for electron that we name the \emph{Schwinger length} of electron\footnote{In all calculations in this paper we drop irrelevant numerical coefficients.}

\be
L_{S}^{d-2} = \frac{\hbar}{\varepsilon_0 c^3}\frac{e^2}{m^2}.
\ee
Obviously this length depends on the particle under consideration. The result \ref{electric-Dvali-Gomez}, though very simple, is very important. It says that the number of electrons inside the Schwinger ball is \emph{proportional to the area of the ball}.
\be
N_e = \frac{R^{d-2}}{L_{S}^{d-2}}.
\ee
This statement, which is the direct result of the inverse area dependence of the Coulomb's law, immediately leads to \emph{Bekenstein entropy} for the Schwinger ball. For this, suppose that the number of degrees of freedom of every quanta (electrons in this case) of the charge ball is $w_0$. Then if we naively suppose that all the entropy of the Schwinger ball comes from the charge ball inside it, then we have
\be
S = N \; ln(w_0) \Rightarrow S \propto \frac{A}{L_{S}^{d-2}},
\ee
where $A$ is the area of the Schwinger ball.

As the final point we note that if we change the charge of the Schwinger ball a little then its area will change a little as well and we have
\be
dQ = \frac{e}{L_S^{d-2}} \; dA.
\ee
As we will see below, in the case of gravity this equation becomes the first law of black hole thermodynamics, since the charge of gravity is mass or energy.

\subsection{Gravitational Schwinger Ball}
We can repeat all previous calculations for gravitational field. A strong enough gravitational field (acceleration\footnote{This acceleration can be considered as a surface gravity in a relativistic notion.}) can create a couple of particles with mass $m$. If we concentrate a lump of massive particles with total mass $M$ in a very small region, then the gravitational field will be stronger than the Schwinger limit inside a region around this mass ball. Previous quantities of the electric Schwinger ball can be converted to their gravitational counterparts by some simple replacements
\bea
&&E_c \rightarrow a_c \cr\cr
&&Q \rightarrow M \cr\cr
&&e \rightarrow m \cr\cr
&&\varepsilon_0 \rightarrow \frac{1}{G},
\eea
where $G$ is the Newton's constant. So the critical acceleration will be
\be\label{critical-acceleration}
a_c =  \frac{mc^3}{\hbar}.
\ee
This critical acceleration can be considered as a gravitaional field, needed to create a couple of particles of mass $m$ in a distance of the order of the de Broglie wavelength of the particles in exact analogy with calculation \ref{Schwinger-simple}.

For massless particles which couple to gravity (including gravitons), any gravitational field can produce a particle with a definite wavelength. There is not a minimum gravitational field to produce such particles\footnote{Free massless particles but not particles in a box. We will shortly see that a black hole is a system of gravitons in a ball.}. For a massless particle with energy $w$, the calculation \ref{Schwinger-simple} is
\bea
\frac{w}{c^2} a \lambda = w \Rightarrow a\lambda = c^2.
\eea
As we will see later (and also we know from Unruh effect), any acceleration $a$ is related to a temperature $T=\hbar a/c$. So the equation above is nothing but the \emph{Wien's displacement law}. It means that we should have a thermal flow of massless particles, produced by acceleration, mostly of wavelength $c^2/a$. But this time the physics behind it, is the Schwinger effect instead of Unruh effect\footnote{In fact the Unruh effect can be considered as a Schwinger effect. For a canonical derivation of the Schwinger effect and more similarities between Schwinger and Unruh effects you can see \cite{Parentani:1991tx}.}.

Now by critical acceleration \ref{critical-acceleration} in hand and by using Newton's law of gravitation\footnote{Here we have assumed that the Newton's law works for strong gravity; at least its form is the same for spherically symmetric sources. This is also verified by general relativity, since the surface gravity of the Schwarzschild solution is $GM/r^{d-2}$. Again any modification to this law will modify our calculations.} the size of the gravitational Schwinger ball obeys
\be\label{G-Dvali-Gomez-1}
R^{d-2} = N \; L_{S}^{d-2},
\ee
where $N = M/m$ is the number of particles inside the ball and the gravitational Schwinger length is
\be
L_{S}^{d-2} = \frac{\hbar G}{c^3}\frac{m^2}{m^2} = \frac{\hbar G}{c^3} = L_P^{d-2},
\ee
which is nothing but the \emph{Planck length}. The important point here is that since the charge of gravity is the mass itself, the gravitational Schwinger length, in contrast to the electric Schwinger length, is a \emph{universal constant}, independent of the particle under consideration. It is a new physics for the Planck length. Now the equation \ref{G-Dvali-Gomez-1} is
\be\label{G-Dvali-Gomez-2}
N = \frac{R^{d-2}}{L_P^{d-2}},
\ee
which is the relation \ref{Dvali-Gomez}, derived by Dvali and Gomez, by borrowing the relation between the mass and radius of the black hole from general relativity. This relation leads to Bekenstein entropy as mentioned in the case of electric Schwinger ball. Remember that here we have not used any relation from general relativity.

Similar to electric Schwinger ball, a small change in the mass of the ball changes the area slightly
\be
dM = m \; d\left( \frac{A}{L_P^{d-2}} \right).
\ee
Now using $m = \hbar \; a_c / c^3 $, it can be rewritten as
\be
dM = \frac{\hbar \; a_c}{c^3} \; d\left( \frac{A}{L_P^{d-2}} \right)
\ee
and this is the first law of black hole thermodynamics with
\bea
&&T = \frac{\hbar \; a_c}{c}\cr\cr
&&S = \frac{A}{L_P^{d-2}},
\eea
which are the Hawking temperature and Bekenstein entropy of the black hole. Note that $a_c$ is the gravitational field at the surface of the Schwinger ball and so comparable to the surface gravity at the horizon of the black hole.

For claiming that the black hole is a gravitational Schwinger ball we need to derive one more relation of the general relativity, which is very important; the relation between the mass and the radius of the Schwarzschild black hole that in general relativity is
\be
R^{d-3}=\frac{G M}{c^2}.
\ee
For this we need to assume that \emph{black hole is a gravitational Schwinger ball made of gravitons that radiates such gravitons}. Graviton ball assumption is also used by Dvali and Gomez in their papers. For a graviton ball we need the replacement below
\be
m \rightarrow \frac{\hbar}{cR}
\ee
and the equation \ref{G-Dvali-Gomez-2} becomes
\bea
&&R^{d-2} = \frac{\hbar G}{c^3}\frac{McR}{\hbar}\cr\cr
\Rightarrow &&R^{d-3} = \frac{GM}{c^2},
\eea
as we require. Now we can claim that \emph{a black hole is a Schwinger ball of gravitons}.

We would like to note that in the picture we drew here, we can have other gravitational Schwinger balls made of other particles but, in a pure theory of gravity, e.g. Einstein-Hilbert action, we just have gravitons.

\section{Schwinger Ball, Black Hole and Information Paradox}

In this section we want to draw a picture of black hole evaporation inspired by the evaporation of the Schwinger ball. We will discuss the information paradox in this picture.

\subsection{Information and Schwinger Ball Evaporation}

As mentioned previousely, inside the Schwinger ball of electrons the electric field is stronger than the Schwinger limit and creates couples of electrons and positrons. The electrons will be pushed away and positrons will be attracted into the Schwinger ball. After interaction with the charge ball at the center of the Schwinger ball, positrons annihilate electrons. Produced photons of this annihilation may escape from the Schwinger ball. The ball will be evaporated completely to electrons and photons.

Now let us prepare the charge ball in a definite pure state initially. From Schwinger's calculation we know that although the created electrons are not thermally distributed but \emph{their spectrum just depends on the strength of the electric field but not the state of the source}. Where is the information of the initial charge ball, when the ball evaporates completely? The answer to this question is so simple and obvious. The information is stored in the \emph{final entangled system of electrons and photons}. But having a look at this evaporation process can be illustrative for black hole evaporation.

A virtual photon from the electric field of the Schwinger ball\footnote{Remember that the classical electric field is in fact a sea of virtual photons.} decays to a couple of electron-positron. Electron (Bob) goes away to infinity and positron (Alice) goes down to the Schwinger ball. Alice and Bob are maximally entangled. Alice interacts with an electron in the charge ball (Carlos) and annihilates it to produce a photon. Now this photon is \emph{entangled with Bob}. The very important point is that the system of photon-Bob includes the information of the \emph{state of Carlos}.

This entangled system of Bob-photon is in fact an EPR couple but, a little \emph{diformed}. Any measurement on Bob (or photon) reduces the state of the system to a definite state that the information of the Carlos can be read from. We name this process of state reduction of the deformed EPR couple as \emph{deformed EPR mechanism}. For example if we find Bob particles thermally distributed after some measurement, then we will find photons non-thermally (or near thermally) distributed to include information of Carlos particles.

One very important point to note is that forgetting photons in the final state (!) will cause many difficulties such as information paradox.

\subsection{Information and Black Hole Evaporation}

Analogous to electric Schwinger ball, we can imagine similar process for black hole evaporation. Since the gravitational field is stronger than the critical value inside the ball, a couple of particles will be created from the effective vaccum (the sea of virtual gravitons)\footnote{Details of this particle production should be determined by the interaction term between gravitons and this hypothetical particle in the correct theory of quantum gravity.}. One of them (Bob) goes to infinity and the other one (Alice) goes into the black hole. Alice interacts with the matter inside (maybe at the center of) the black hole, and annihilates a particle (Carlos) to produce a graviton (Grace). This annihilation event is related to the initial pair production by a \emph{time reversal transformation}. This system of Bob-Grace is a deformed EPR couple which also includes the information of the state of Carlos.

There are some important differences between gravitational and electric Schwinger balls. The first important difference is that \emph{gravitons couple to gravitons}. This causes more complexity in the process. The other important difference is that gravitons are trapped into the Schwinger ball because of interaction between gravitons. So the produced gravitons can not escape from the ball. This entrapment has some important consequences.

\begin{figure}
\centering
\includegraphics[scale=.8]{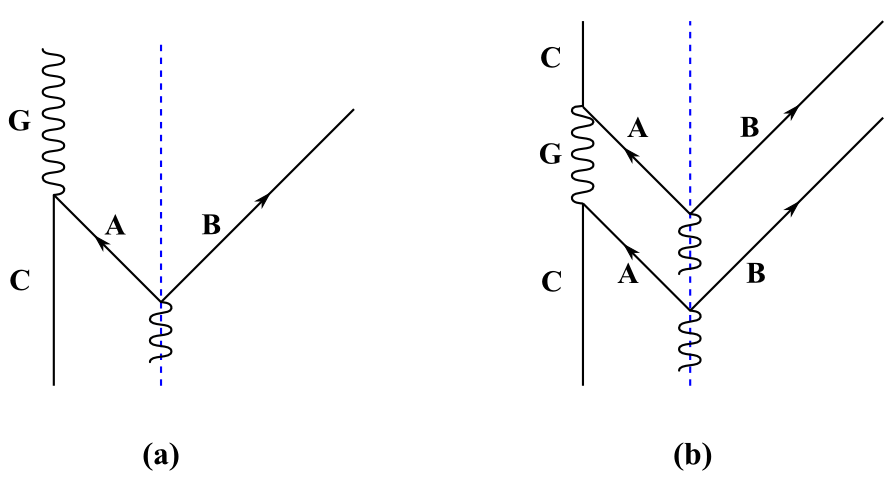}
\caption{\label{revolutions}A sketch of the Feynman diagrams of events which lead to first and second revolutions. (a) An incoming Alice annihilates Carlos to produce a Grace which is entangled with the Bob. (b) An incoming Alice annihilates a previousely produced Grace to a new Carlos which is entangled with both old and young Bobs.}
\end{figure}

Let us think about an extremly simplified model of evaporation. Initialy we have Carlos particles at the center of a gravitational Schwinger ball surrounded by an ultra critical gravitational field (sea of gravitons). This sea of gravitons creates couples of Alice and Bob. Alice interacts with Carlos and produce a Grace which is trapped in the ball. Suppose that (and just suppose that!) every incoming Alice annihilates a Carlos as long as Carlos particles exist\footnote{Every incoming Alice is free to interact with a Carlos or a Grace. In fact when there are more Carlos particles than Grace particles, Alice more likely interacts with a Carlos. When the number of Carlos and Grace particles become the same, Alice interacts with them equally. But, here we suppose the explained process for more clarity.}. Since Graces are trapped in the ball, every annihilation event increases the entanglement entropy of the black hole and Bobs. The maximum entanglement will be achieved when all Carlos particles are annihilated. Let us name this state \emph{the first revolution} which there is not any Carlos particle inside the black hole anymore. After this moment any incoming Alice, which is entangled to its Bob, has to interact with a Grace to produce a new Carlos. The new Carlos is simultaneously entangled to early (old) Bob and late (young) Bob. This simultaneous entanglement has the same von Neumann entropy of the couple of old Bob-Grace. Again suppose that (and just suppose that!), after first revolution, any incoming Alice interacts with a Grace to produce a new Carlos until all Graces are annihilated. Then we will have a system of new Carlos particles, inside the ball, each one entangled with a couple of old-young Bob particles out of the ball. This state that we name \emph{the second revolution}, has the same entanglement entropy of the first revolution. The first and second revolutions are schematically sketched in fig \ref{revolutions}.

The system experiences the third, fourth and more revolutions until the mass of the black hole becomes so small that can not create particles inside the ball. This happens because even massless particles have a lowest wavelength equal to the size of the ball. So the system can not be a Schwinger ball and there will not be any black hole finally. An exact explanation of the last steps of the evaporation process probably needs to have a quantum theory of gravity.

\subsection{More Realistic Evaporation Process}

In previous part we explained an extremely simplified process of evaporation just for clarity. We need to consider more realistic situations. As mentioned previousely, when the ball starts to evaporate, incoming Alices can interact with both Carlos and Grace particles. The number of Grace particles increases and the number of Carlos particles decreases because of these interactions. The numbers of different types of particles should tend to nearly equilibrated ratios, as time goes, although the total number and energy of the particles decrease. On the other hand Grace and primary (or secondary) Carlos particles \emph{can merge} and produce new Alice particles with much more complicated entanglement patterns with Bobs. It seems obvious that at some moment the entanglement between the systems inside and outside the ball becomes maximum. In our previous simplified picture this happens at the moment of first revolution and in a more realistic case it occures when the number of Grace and Carlos particles becomes the same and we can also imagine more complicated situations.

As is shown in the appendix, the simple annihilations by Alices do not change the entanglement entropy of the black hole after the Page's time. The entanglement entropy decreases only when the number of particles, inside the ball, decreases. This happens by \emph{merging} and \emph{tunneling out}.

After the Page's time the entanglement entropy of the black hole will decrease because of mergings inside the ball. These mergings produce new particles inside the black hole, entangled with more and more Bob particles with more and more complicated entanglement patterns. Every merging event reduces the entanglement entropy by one unit\footnote{For example for simple spin 1/2 states this unit is $ln2$.}.

On the other hand, during evaporation, some particles may escape from the ball by quantum tunneling. Specially this tunneling may be more important at final moments of the evaporation process.

These merging and tunneling events seem inevitable for gravitons inside the black hole. Very soft gravitons can not live in the ball when its radius decreases. They have to escape from the ball or merge and produce more energetic gravitons. Both these events reduce the entanglement entropy of the black hole.

When evaporation is completed, the final system is made of a hugh number of entangled particles in a pure state. The evolution is unitary and information of the initial system, made of primary Carlos particles, is encoded in the final system. Again forgetting the role of interaction between Carlos and Alice particles and also the role of gravitons can cause many difficulties such as information paradox. In the next part we explain why firewall paradox arises and how will be resolved in the process we explained above.

\subsection{Firewall Problem}

The famous firewall problem \cite{Almheiri:2012rt} states that the three statements of equivalence principle, locality and purity of the final state which is related to unitarity of the evaporation process, are inconsistent and we need to give up one of them. The authors then give up the equivalence principle since locality and unitarity seem to be more fundamental. Here we draw a completely different picture of the process.

The mixture of equivalence principle and local quantum field theory says that the couples of Alice and Bob can be removed by a coordinate transformation. It means that their effects should be removable as well. If we want to remove these couples and all their effects completely by just a coordinate transformation, they should be disentangled from the rest of the universe. So the entanglement entropy of the system of Alice-Bob with the rest of the universe should be zero.
\be
S_{Alice-Bob} = 0.
\ee
On the other hand if we want the final state to be pure, we need some entanglement between young and old Bobs after the Page's time and these two requirements are inconsistent.\emph{ It is not possible to have the couple of Alice-young Bob, disentangled from the rest of the universe, and simultaneously the entangled couple of young-old Bobs}. This paradox can be shown by the use of strong subaditivity.
\be\label{strong-subadditivity}
S_{Alice-young}+S_{young-old} \geq S_{young}+S_{Alice-young-old}.
\ee
Unitarity and purity of the final system need the entanglement entropy of the radiation to be decreasing after the Page's time \cite{Page:1993wv}
\be\label{firewall1}
S_{young-old} < S_{old}.
\ee
On the other hand disentanglement of Alice-Bob from the rest of the universe implies
\bea\label{firewall2}
&&S_{Alice-young} = 0\cr\cr
&&S_{Alice-young-old} = S_{old}.
\eea
But using equations \ref{firewall1} and \ref{firewall2} in \ref{strong-subadditivity} leads to
\be
S_{young} < 0,
\ee
which is impossible. For solving this inconsistency authors of \cite{Almheiri:2012rt} relax the equations \ref{firewall2} which come from disentanglement of Alice-Bob from the rest of the universe implied by the equivalence principle.

\begin{figure}
\centering
\includegraphics[scale=.8]{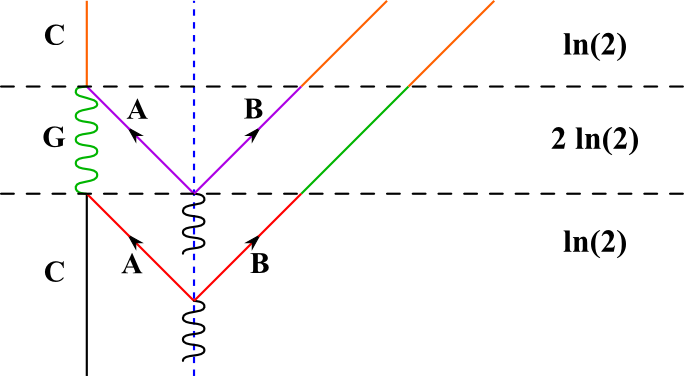}
\caption{\label{firewall}A sketch of evaporation process with corresponding entanglement entropy of the black hole if the particles are just spin states. Different entangled particles are shown by the same colors. When the purple couple of A-B is created, the entanglement entropy is increased. After annihilation the entanglement entropy returns back to its initial value.}
\end{figure}

Now let us have a careful look at the evaporation process with the mechanism that we explained in the previous part. After the first revolution (Page's time) the created couple of Alice-Bob, before annihilation, is disentangled from the rest of the universe as the equivalence principle demands. But exactly at the same time the equation \ref{firewall1} is not correct and in fact we have
\be
S_{old-young} = S_{old}+S_{young} > S_{old}.
\ee
But how does entanglement entropy of the black hole with the rest of the universe decreases in order to have a final pure state as Page has shown?

As mentioned previously, when Alice annihilates Grace the entanglement entropy between the new Carlos and old-young Bobs is the same as the entanglement entropy between Grace and old Bob before annihilation. The entanglement entropy decreases when the number of particles inside the black hole decreases, by merging or tunneling out.

After annihilation the couple of Alice-Bob does not exist anymore. Instead we have the system of (new Carlos)-(old Bob)-(young Bob) in a pure state. Breaking this system in any way is consistent with strong subadditivity. Whole this process is shown if fig. \ref{firewall}.

On the other hand one may consider an escaped particle (by tunneling from the ball which reduces the entanglement entropy) as a young Bob which is obviously entangled with some old Bobs. In this case this escaped Bob does not have an Alice partner so $S_{Alice-young}=S_{young}$ and $S_{Alice-young-old}=S_{young-old}$ and the strong subadditivity relation \ref{strong-subadditivity} becomes trivial.

The equations \ref{firewall1} and \ref{firewall2} were inconsistent if they were the results of the same event but, they are not.

\section{Summary and Discussion}
In this paper we showed that the inverse area dependence of Coulomb's and Newton's laws, when coupled to Schwinger effect, results the Bekenstein entropy for electric and gravitational Schwinger balls. Interestingly since the charge of gravity is mass (energy), the gravitational Schwinger length is a universal constant which is nothing but the Planck length. Also a small change in the mass of the gravitational Schwinger ball leads to the first law of black hole thermodynamics. The relation between the mass and the size of the Schwarzschild black hole can also be derived if the Schwinger ball is made of gravitons. All these characteristics suggest that the Schwarzschild black hole is a gravitational Schwinger ball of gravitons.

Inspired by electric Schwinger ball we drew a microscopic picture of black hole evaporation which can resolve both the information and firewall paradoxes. This microscopic process of evaporation implies that the entanglement entropy of the black hole reduces just when the number of particles inside the black hole decreases, e.g. by merging or tunneling out.

In this picture, inside the black hole as a Schwinger ball is a regular region of space, full of gravitons. Spatial and temporal coordinates do not change role and the horizon is not a coordinate singularity and also there is not a singularity at the center of the black hole like what general relativity results. All these strange features can be some fake by-products of the effective and emergent concept of the curved space of general relativity. If black holes are really Schwinger balls, then it means that there is something wrong with general relativity or it is not applicable to such strong gravitational fields. But these strange characteristics of black holes in general relativity show that something special is happening at the horizon and inside the black hole.

The observer dependency of the particle production in Hawking and Unruh effects, in the language of Schwinger effect, comes from the observer dependency of the acceleration. Different non-inertial observers experience different accelerations or gravitational fields. It leads to different particle production rates for different observers. Suppose that we have an observer in a spaceship, hovering far away from a black hole. He uses engines of his spaceship to cancel the gravitational attraction of the black hole. He experiences a gravitational attraction $a_0$. If he lowers down the power of the engines, then he will start to fall towards the black hole, but not freely. Now he experiences the attraction $a_1<a_0$. So he sees a cooler black hole than before. If he turns off the engines then he will not see any radiating horizon when he reaches it. In fact by lowering down the power of engines, the horizon fades out, from the viewpoint of the observer. This happens since the (gravitational attractive) acceleration is observer dependent and so the Schwinger effect is observer dependent as well. Of course since this comes from the equivalence principle, any modification to this principle from quantum gravity will change this picture.

Note that the Schwinger ball picture of the black hole implies that although the free falling observer does not see any particle production around himself but, he sees that the lower layers of the space radiate because of the gradient of the gravitational field.

As the final point we want to propose that the phase transition of a system of gravitons, proposed by Dvali and Gomez, can be related to the gravitational Schwinger effect that happens when the graviton ball becomes small enough. Having firm statements on this needs more investigations.

\section{Appendix}
Suppose that a Carlos particle with spin state $\vert\uparrow\rangle $ is annihilated by an Alice particle which is entangled with a Bob partner in a singlet state. This annihilation can be shown by transition below
\be
\vert\uparrow\rangle \otimes \frac{1}{\sqrt{2}}\bigg[ |\uparrow \rangle \otimes |\downarrow \rangle  - |\downarrow \rangle \otimes |\uparrow \rangle \bigg] \rightarrow
 P_1 |1 \rangle \otimes |\downarrow \rangle  + P_2 |0\rangle \otimes |\uparrow \rangle .
\ee
$P_1$ and $P_2$ are two coefficients, satisfy $|P_1|^2+|P_2|^2 = 1$. The values of these coefficients depend on details of the interaction between particles in microscopic theory, e.g. quantum theory of gravity. It is easy to show that the reduced density matrix is
\be
\rho _{BH} = |P_1|^2 \; |\downarrow \rangle \langle \downarrow | + |P_2|^2 \; |\uparrow \rangle \langle \uparrow |,
\ee
so the entanglement entropy of the black hole will be
\be
S_{BH}=-|P_1|^2\;ln|P_1|^2 - |P_2|^2\; ln |P_2|^2.
\ee
Particles are maximally entangled when $P_1=P_2=1/\sqrt{2}$. Now suppose that another couple of Alice and Bob is created. Before annihilation the state of the system is
\be
\bigg[ P_1\;|1 \rangle \otimes |\downarrow \rangle + P_2\;|0 \rangle \otimes |\uparrow \rangle \bigg]\otimes \frac{1}{\sqrt{2}}\bigg[ |\uparrow \rangle \otimes |\downarrow \rangle  - |\downarrow \rangle \otimes |\uparrow \rangle \bigg].
\ee
Again it is trivial that the reduced density matrix is
\bea
\rho _{BH}=\frac{1}{2}\bigg[ &&|P_1|^2\;|\downarrow \rangle \langle \downarrow |\otimes |\downarrow \rangle \langle \downarrow | + |P_1|^2\;|\downarrow \rangle \langle \downarrow |\otimes |\uparrow \rangle \langle \uparrow |\cr\cr
+ &&|P_2|^2\;|\uparrow \rangle \langle \uparrow |\otimes |\downarrow \rangle \langle \downarrow |+|P_2|^2\;|\uparrow \rangle \langle \uparrow |\otimes|\uparrow \rangle \langle \uparrow |\bigg].
\eea
So the black hole entanglement entropy, before annihilation, will be
\be
S_{BH} = ln2 - |P_1|^2\;ln|P_1|^2 - |P_2|^2\; ln |P_2|^2.
\ee
Now we calculate the black hole entanglement entropy after annihilation. When Alice annihilates Grace to produce a new Carlos, we have the transition below
\bea
&&\frac{1}{\sqrt{2}}\bigg[ P_1\;|1 \rangle \otimes |\downarrow \rangle  + P_2\;|0 \rangle \otimes |\uparrow \rangle \bigg]\otimes \bigg[ |\uparrow \rangle \otimes |\downarrow \rangle  - |\downarrow \rangle \otimes |\uparrow \rangle \bigg]\cr\cr
\rightarrow && \hat{P}_1\;|\uparrow \rangle _{BH} \otimes |\uparrow \rangle \otimes |\downarrow \rangle + \hat{P}_2\;|\uparrow \rangle _{BH}  \otimes |\downarrow \rangle \otimes |\uparrow \rangle + \hat{P}_3\; |\downarrow \rangle _{BH}  \otimes |\uparrow \rangle \otimes |\uparrow \rangle .
\eea
$\hat{P}_1$, $\hat{P}_2$ and $\hat{P}_3$ are new coefficients with the condition $|\hat{P}_1|^2+|\hat{P}_2|^2+|\hat{P}_3|^2 = 1$. The subscript $BH$ refers to the particle inside the black hole. Note that in the state above we do not have the state $|3/2 \rangle \otimes |\downarrow \rangle \otimes |\downarrow \rangle$ since there is not a particle with spin $3/2$ in the spectrum of our theory. This is the case for any other theory since there is always a particle with highest spin.

For calculating the entanglement entropy of the black hole, we should trace the state of the black hole out and then calculate the von Neumann entropy of the reduced density matrix. Instead we do a simpler work. Since the total state is pure, we can trace out the state of particles outside the black hole, with the same result. For this we can rewrite the state above in this form
\bea
&&\sqrt{|\hat{P}_1|^2+|\hat{P}_2|^2}\;\bigg[\frac{\hat{P}_1}{\sqrt{|\hat{P}_1|^2+|\hat{P}_2|^2}}\;|\uparrow \rangle \otimes |\downarrow \rangle + \frac{\hat{P}_2}{\sqrt{|\hat{P}_1|^2+|\hat{P}_2|^2}}\;|\downarrow \rangle \otimes |\uparrow \rangle\bigg]\otimes |\uparrow \rangle _{BH}\cr\cr
+ &&\hat{P}_3\; |\uparrow \rangle \otimes |\uparrow \rangle \otimes |\downarrow \rangle _{BH}.
\eea
Now you can check that the reduced density matrix is
\bea
\rho_{out} = (|\hat{P}_1|^2+|\hat{P}_2|^2)\; |\uparrow \rangle \langle \uparrow | + |\hat{P}_3|^2 \; |\downarrow \rangle \langle \downarrow |,
\eea
which leads to
\be
S_{out}=S_{BH}= - \big(|\hat{P}_1|^2+|\hat{P}_2|^2\big)\;ln\big(|\hat{P}_1|^2+|\hat{P}_2|^2\big) - |\hat{P}_3|^2\;ln|\hat{P}_3|^2.
\ee

The result above is general, when a particle inside the black hole is entangled with $n$ particles outside the black hole in a total spin up (down) state. To show this remember that the state of such system is
\be
|\psi \rangle = P_1\;|\uparrow \rangle _{BH} \otimes |\phi\rangle + P_2\;|\downarrow \rangle _{BH} \otimes |\chi\rangle ,
\ee
where $|\phi\rangle$ is a normalized state with $S_z = 0$, made of $n$ spin 1/2 states and $|\chi\rangle$ is a state with $S_z = 1$, so perpendicular to $|\phi\rangle$.
Since the total state is pure, we can trace out the particles outside instead of the particle inside the black hole. The resulting reduced density matrix is
\be
\rho _{out} = |P_1|^2\; |\uparrow \rangle \langle \uparrow | + |P_2|^2 \; |\downarrow \rangle \langle \downarrow |,
\ee
with entropy
\be
S_{out}=S_{BH}= - |P_1|^2\;ln|P_1|^2 - |P_2|^2\;ln|P_2|^2.
\ee
Again it means that when the interaction term does not prefer one spin state, and so $P_1=P_2$, independent of $n$, increasing the number of particles outside the black hole which are entangled with one particle inside the black hole, does not change the entanglement entropy. The only way to decrease the entanglement entropy is reducing the number of particles inside the black hole by merging or tunneling out.

\vspace{5mm}
\hrule



\begin{thebibliography}{99}


\bibitem{Bardeen:1973gs} 
  J.~M.~Bardeen, B.~Carter and S.~W.~Hawking,
  ``The Four laws of black hole mechanics,''
  Commun.\ Math.\ Phys.\  {\bf 31}, 161 (1973).
  doi:10.1007/BF01645742
 
\bibitem{Hawking:1974sw} 
  S.~W.~Hawking,
  ``Particle Creation by Black Holes,''
  Commun.\ Math.\ Phys.\  {\bf 43}, 199 (1975)
  Erratum: [Commun.\ Math.\ Phys.\  {\bf 46}, 206 (1976)].
  doi:10.1007/BF02345020


\bibitem{Hawking:1976ra} 
  S.~W.~Hawking,
  ``Breakdown of Predictability in Gravitational Collapse,''
  Phys.\ Rev.\ D {\bf 14}, 2460 (1976).
  doi:10.1103/PhysRevD.14.2460


\bibitem{Mathur:2009hf} 
  S.~D.~Mathur,
  ``The Information paradox: A Pedagogical introduction,''
  Class.\ Quant.\ Grav.\  {\bf 26}, 224001 (2009)
  doi:10.1088/0264-9381/26/22/224001
  [arXiv:0909.1038 [hep-th]].


\bibitem{Dvali:2015aja} 
  G.~Dvali,
  ``Non-Thermal Corrections to Hawking Radiation Versus the Information Paradox,''
  Fortsch.\ Phys.\  {\bf 64}, 106 (2016)
  doi:10.1002/prop.201500096
  [arXiv:1509.04645 [hep-th]].


\bibitem{'tHooft:1990fr} 
  G.~'t Hooft,
  ``The black hole interpretation of string theory,''
  Nucl.\ Phys.\ B {\bf 335}, 138 (1990).
  doi:10.1016/0550-3213(90)90174-C

\bibitem{Susskind:1993if} 
  L.~Susskind, L.~Thorlacius and J.~Uglum,
  ``The Stretched horizon and black hole complementarity,''
  Phys.\ Rev.\ D {\bf 48}, 3743 (1993)
  doi:10.1103/PhysRevD.48.3743
  [hep-th/9306069].

\bibitem{Giddings:2012gc} 
  S.~B.~Giddings,
  ``Nonviolent nonlocality,''
  Phys.\ Rev.\ D {\bf 88}, 064023 (2013)
  doi:10.1103/PhysRevD.88.064023
  [arXiv:1211.7070 [hep-th]].


\bibitem{Frolov:1993jq} 
  V.~P.~Frolov and I.~Novikov,
  ``Wormhole as a device for study black hole's interior,''
  Phys.\ Rev.\ D {\bf 48}, 1607 (1993).
  doi:10.1103/PhysRevD.48.1607

\bibitem{Page:1993up} 
  D.~N.~Page,
  ``Black hole information,''
  hep-th/9305040.

\bibitem{Maldacena:2013xja} 
  J.~Maldacena and L.~Susskind,
  ``Cool horizons for entangled black holes,''
  Fortsch.\ Phys.\  {\bf 61}, 781 (2013)
  doi:10.1002/prop.201300020
  [arXiv:1306.0533 [hep-th]].

 \bibitem{Dvali-Gomez-papers}
G.~Dvali and C.~Gomez, ``Black Hole's Quantum N-Portrait", Fortsch.Phys. 61 (2013) 742-767,  
 arXiv:1112.3359 [hep-th]. 
 \\
 ``Black Hole's 1/N Hair,''
 Phys.\ Lett.\ B {\bf 719} (2013) 419,  [arXiv:1203.6575 [hep-th]];  
  \\
   ``Black Holes as Critical Point of Quantum Phase Transition" 
 Eur.Phys.J. C74 (2014) 2752,  arXiv:1207.4059 [hep-th];   
 
 ~\\
 	``Black Hole Macro-Quantumness",  arXiv:1212.0765 [hep-th];
 
 

\bibitem{Fulling:1972md} 
  S.~A.~Fulling,
  ``Nonuniqueness of canonical field quantization in Riemannian space-time,''
  Phys.\ Rev.\ D {\bf 7}, 2850 (1973).
  doi:10.1103/PhysRevD.7.2850

\bibitem{Davies:1974th} 
  P.~C.~W.~Davies,
  ``Scalar particle production in Schwarzschild and Rindler metrics,''
  J.\ Phys.\ A {\bf 8}, 609 (1975).
  doi:10.1088/0305-4470/8/4/022

\bibitem{Unruh:1976db} 
  W.~G.~Unruh,
  ``Notes on black hole evaporation,''
  Phys.\ Rev.\ D {\bf 14}, 870 (1976).
  doi:10.1103/PhysRevD.14.870

\bibitem{Sauter:1931zz} 
  F.~Sauter,
  ``Uber das Verhalten eines Elektrons im homogenen elektrischen Feld nach der relativistischen Theorie Diracs,''
  Z.\ Phys.\  {\bf 69}, 742 (1931).
  doi:10.1007/BF01339461

\bibitem{Heisenberg:1935qt} 
  W.~Heisenberg and H.~Euler,
  ``Consequences of Dirac's theory of positrons,''
  Z.\ Phys.\  {\bf 98}, 714 (1936)
  doi:10.1007/BF01343663
  [physics/0605038].

\bibitem{Schwinger:1951nm} 
  J.~S.~Schwinger,
  ``On gauge invariance and vacuum polarization,''
  Phys.\ Rev.\  {\bf 82}, 664 (1951).
  doi:10.1103/PhysRev.82.664


\bibitem{Parentani:1991tx} 
  R.~Parentani and R.~Brout,
  ``Vacuum instability and black hole evaporation,''
  Nucl.\ Phys.\ B {\bf 388}, 474 (1992).
  doi:10.1016/0550-3213(92)90623-J
  
  
\bibitem{Frolov:2014wja} 
  V.~P.~Frolov,
  ``Do Black Holes Exist?,''
  arXiv:1411.6981 [hep-th].
  
\bibitem{Kim:2016dmm} 
  S.~P.~Kim,
  ``Schwinger Effect, Hawking Radiation, and Unruh Effect,''
  arXiv:1602.05336 [hep-th].


\bibitem{Almheiri:2012rt} 
  A.~Almheiri, D.~Marolf, J.~Polchinski and J.~Sully,
  ``Black Holes: Complementarity or Firewalls?,''
  JHEP {\bf 1302}, 062 (2013)
  doi:10.1007/JHEP02(2013)062
  [arXiv:1207.3123 [hep-th]].


\bibitem{Page:1993wv} 
  D.~N.~Page,
  ``Information in black hole radiation,''
  Phys.\ Rev.\ Lett.\  {\bf 71}, 3743 (1993)
  doi:10.1103/PhysRevLett.71.3743
  [hep-th/9306083].






\end{thebibliography}
\end{document}